\documentclass[twocolumn,trackchanges]{aastex701}
\usepackage{amsmath}

\shorttitle{SN 2025wiu}
\shortauthors{Lai et al.}
\graphicspath{{./}}

\begin{document}

\title{SN~2025wiu: A Type Ib Supernova with Post-Peak Optical Rebrightening}

\author[0009-0006-7037-0594]{Cheng-Han~Lai}
\affiliation{Graduate Institute of Astronomy, National Central University, 300 Jhongda Road, 32001 Jhongli, Taiwan}
\email{m1129008@astro.ncu.edu.tw}

\author[0000-0002-1066-6098]{Ting-Wan~Chen}
\affiliation{Graduate Institute of Astronomy, National Central University, 300 Jhongda Road, 32001 Jhongli, Taiwan}
\email{twchen@astro.ncu.edu.tw}

\author[0009-0000-2423-6825]{Yu-Hao~Zhang}
\affiliation{Institute of Astrophysics, Central China Normal University, Wuhan 430079, China}
\email{zhang-yh@mails.ccnu.edu.cn}

\author[0000-0002-8708-0597]{Liang-Duan~Liu}
\affiliation{Institute of Astrophysics, Central China Normal University, Wuhan 430079, China}
\email{liuld@ccnu.edu.cn}

\author[0000-0003-1546-6615]{Jesper~Sollerman}
\affiliation{The Oskar Klein Centre, Department of Astronomy, Stockholm University, AlbaNova, SE-10691 Stockholm, Sweden}
\email{jesper@astro.su.se}

\author[0000-0002-9928-0369]{Amar~Aryan}
\affiliation{Graduate Institute of Astronomy, National Central University, 300 Jhongda Road, 32001 Jhongli, Taiwan}
\email{amar@astro.ncu.edu.tw}

\author[0000-0002-7708-3831]{Anirban~Dutta}
\affiliation{Graduate Institute of Astronomy, National Central University, 300 Jhongda Road, 32001 Jhongli, Taiwan}
\email{anirbaniamdutta@gmail.com}

\author[0000-0002-2898-6532]{Sheng~Yang}
\affiliation{Institute for Gravitational Wave Astronomy, Henan Academy of Sciences, Zhengzhou 450046, Henan, China}
\email{sheng.yang@hnas.ac.cn}

\author[0000-0002-4223-103X]{Christoffer~Fremling}
\affiliation{Caltech Optical Observatories, California Institute of Technology, Pasadena, CA 91125, USA}
\affiliation{Division of Physics, Mathematics, and Astronomy, California Institute of Technology, Pasadena, CA 91125, USA}
\email{fremling@caltech.edu}

\author[0000-0002-3168-0139]{Matthew~J.~Graham}
\affiliation{Division of Physics, Mathematics, and Astronomy, California Institute of Technology, Pasadena, CA 91125, USA}
\email{mjg@caltech.edu}

\author[]{David~Hale}
\affiliation{Caltech Optical Observatories, California Institute of Technology, Pasadena, CA 91125, USA}
\email{dhale@caltech.edu}

\author[0000-0002-0129-806X]{K.-Ryan Hinds}
\affiliation{Division of Physics, Mathematics, and Astronomy, California Institute of Technology, Pasadena, CA 91125, USA}
\email{khinds@caltech.edu}

\author[0000-0002-5619-4938]{Mansi~M.~Kasliwal}
\affiliation{Division of Physics, Mathematics, and Astronomy, California Institute of Technology, Pasadena, CA 91125, USA}
\email{mansi@astro.caltech.edu}

\author[0000-0002-8532-9395]{Frank~J.~Masci}
\affiliation{IPAC, California Institute of Technology, 1200 E. California Blvd, Pasadena, CA 91125, USA}
\email{fmasci@ipac.caltech.edu}

\author[0000-0002-0387-370X]{Reed~L.~Riddle}
\affiliation{Caltech Optical Observatories, California Institute of Technology, Pasadena, CA 91125, USA}
\email{riddle@caltech.edu}

\author[0000-0003-4725-4481]{Sam~Rose}
\affiliation{Division of Physics, Mathematics, and Astronomy, California Institute of Technology, Pasadena, CA 91125, USA}
\email{srose@caltech.edu}

\author[0000-0001-7648-4142]{Ben~Rusholme}
\affiliation{IPAC, California Institute of Technology, 1200 E. California Blvd, Pasadena, CA 91125, USA}
\email{rusholme@ipac.caltech.edu}

\author[]{Niharika~Sravan}
\affiliation{Department of Physics, Drexel University, Philadelphia, PA 19104, USA}
\email{niharika.sravan@gmail.com}

\correspondingauthor{Cheng-Han~Lai}
\email[show]{m1129008@astro.ncu.edu.tw}  

\correspondingauthor{Ting-Wan~Chen}
\email[show]{twchen@astro.ncu.edu.tw}  

\correspondingauthor{Yu-Hao~Zhang}
\email[show]{zhang-yh@mails.ccnu.edu.cn}

\begin{abstract}
We present optical photometric and optical/near-infrared spectroscopic observations of SN~2025wiu, a Type Ib supernova (SN) at $z=0.046927$. The $r$-band light curve exhibits an initial maximum of approximately $-18.2$ mag, followed by a weaker optical rebrightening reaching approximately $-17.8$ mag at $+38$ days in the rest frame relative to the first maximum. However, the reconstructed bolometric light curve does not show a distinct second peak, but instead exhibits a reduced decline rate during the corresponding phase. The spectra consistently show Type Ib features without narrow emission components associated with circumstellar material. The late-time bolometric decline rate of $0.054\pm0.008$ mag day$^{-1}$ is faster than expected for complete trapping of $^{56}$Co decay. A radioactive-decay model for the first peak requires $M_{\rm Ni}=0.47~M_\odot$, above the range typical of SNe Ib. A combined radioactive-decay and circumstellar-interaction model reproduces the full bolometric light curve with $M_{\rm Ni}\approx0.38~M_\odot$, $M_{\rm ej}\approx4.7~M_\odot$, and an effective circumstellar mass of $\approx0.10~M_\odot$, with the interaction contributing from the first peak onward. The lack of interaction signatures in the spectra favours an aspherical, disk-like circumstellar structure engulfed by the ejecta, in which the interaction energy is thermalized below the ejecta photosphere; the fitted circumstellar parameters are therefore effective values rather than direct measurements of the circumstellar geometry. SN~2025wiu demonstrates that optical rebrightening does not necessarily imply a genuine bolometric secondary maximum and highlights the importance of multi-band observations for identifying the powering mechanisms of double-peaked stripped-envelope supernovae.
\end{abstract}


\section{Introduction} \label{sec:Introduction}

Massive stars with $M \gtrsim 8\,M_\odot$ end their lives as core-collapse supernovae \citep[CCSNe; e.g.,][]{2002RvMP...74.1015W,2003ApJ...591..288H,2009ARA&A..47...63S}.Those that retain their hydrogen-rich envelopes show prominent hydrogen features in their spectra and are classified as Type~II, whereas stripped-envelope supernovae (SESNe) arise from progenitors that have lost their outer hydrogen layers prior to explosion and therefore lack hydrogen features. Several mechanisms have been proposed for this stripping, including strong stellar winds from massive single stars \citep{1975MSRSL...9..193C,2014ARA&A..52..487S} and mass transfer or common-envelope evolution in binary systems \citep{1992ApJ...391..246P,2010ApJ...725..940Y,2011MNRAS.412.1522S}. SESNe are further subdivided according to the degree of stripping: objects that retain their helium layers are classified as Type~Ib, while those in which helium has also been removed are classified as Type~Ic \citep{1997ARA&A..35..309F}.

The optical emission of SESNe is predominantly powered by the radioactive decay chain $^{56}$Ni~$\rightarrow$~$^{56}$Co~$\rightarrow$~$^{56}$Fe, in which the gamma rays released by the decay are thermalized in the expanding ejecta \citep{1962PhDT........25P,1969ApJ...157..623C,1982ApJ...253..785A}. This mechanism produces a single, smooth light curve that peaks when the diffusion timescale becomes comparable to the expansion timescale. Events that are more luminous or longer-lived than can be accommodated by physically plausible $^{56}$Ni masses require additional power sources, most commonly the spin-down of a rapidly rotating, highly magnetized neutron star \citep{2010ApJ...717..245K,2010ApJ...719L.204W}, or the conversion of kinetic energy into radiation through the collision between the ejecta and circumstellar material (CSM) lost by the progenitor prior to explosion \citep{1994ApJ...420..268C,2012ApJ...746..121C,2013MNRAS.435.1520M}. When the CSM is dense ($\dot{M} \sim 10^{-3}$--$1\,M_\odot\,\mathrm{yr^{-1}}$; \citealt{2012ApJ...744...10K,2014MNRAS.439.2917M,2025ApJ...987...13R}) and close to the progenitor ($r \sim 10^{14}$--$10^{16}$\,cm; \citealt{2014MNRAS.439.2917M,2022ApJ...927...25M}), the interaction produces characteristic narrow emission lines, giving rise to the Type~IIn, Ibn, and Icn classes for hydrogen-, helium-, and carbon/oxygen-rich CSM, respectively \citep{1990MNRAS.244..269S,2017ApJ...836..158H,2022Natur.601..201G,2022ApJ...927..180P}.

While typical SESNe show a single-peaked light curve followed by a smooth decline onto the radioactive tail, double-peaked SESNe show a rebrightening after the first peak, which is difficult to explain with $^{56}$Ni decay alone; \citet{2025PASP..137i4101S} provide a systematic comparison of the possible mechanisms. Interaction with CSM offers considerable flexibility: depending on the density profile, geometry, and location of the material, it can produce a wide variety of light-curve morphologies \citep{2024ApJ...973...14C}, ranging from a single luminous bump \citep{2023A&A...678A.209K} to multiple undulations arising from a clumpy CSM distribution \citep{2026A&A...706A.183P}. Delayed energy injection from a magnetar has likewise been invoked, either through variations in the thermalization of the spin-down energy \citep{2022MNRAS.513.6210M}, through evaporation of a companion star \citep{2024ApJ...970L..42Z}, or through magnetar flaring activity \citep{2023ApJ...951...61D}. A third possibility is a double-peaked $^{56}$Ni distribution, in which nickel is present both in the inner and outer ejecta. Because this scenario invokes no additional energy source, the total luminosity is bounded by the synthesized nickel mass and the separation between the two peaks is tightly constrained by the spatial distribution of the nickel, restricting the range of light-curve morphologies it can reproduce; SN\,2005bf \citep{2006ApJ...641.1039F}, PTF11mnb \citep{2018A&A...609A.106T} and SN\,2019cad \citep{2022A&A...667A..92O} have been interpreted in this way. Other mechanisms, including fallback accretion onto the compact remnant \citep{2023ApJ...956L..31M, 2024Natur.625..253C, 2025A&A...700L..17O,2026A&A...707A.161C} and pulsational pair instability \citep{2024ApJ...977L..41A}, have also been proposed.

In this paper, we present SN\,2025wiu, a double-peaked Type~Ib supernova (SN Ib) at $z = 0.047$. Its well-sampled Zwicky Transient Facility \citep[ZTF;][]{2019PASP..131a8002B,2019PASP..131g8001G,2020PASP..132c8001D,masci2019zwicky} and Asteroid Terrestrial-impact Last Alert System \citep[ATLAS;][]{tonry2018atlas,smith2020design} light curves show a second peak $\sim$38~days after the first, followed by a decline considerably faster than expected from $^{56}$Co decay with full $\gamma$-ray trapping. Our 11 spectra, including a near-infrared spectrum obtained near the second peak, retain Type~Ib characteristics throughout, with no narrow emission lines at any epoch and essentially no change across the second peak. The $g-r$ color shows no significant blueward evolution during the second peak, and the bolometric light curve exhibits no distinct second maximum. With a peak absolute magnitude of $M_r \approx -18.2$~mag, SN\,2025wiu is also considerably less luminous than SN\,2019stc \citep{2021ApJ...913..143G} and SN\,2021uvy \citep{2025PASP..137i4101S}. This combination of properties makes SN\,2025wiu a valuable test case for interaction-powered rebrightening without spectroscopic signatures, and we argue that its second peak arises from ongoing interaction between the ejecta and an embedded, disk-like CSM.

This paper is organized as follows. In Section~\ref{sec:Observations}, we present our observations and data reduction. In Section~\ref{sec:Photometric}, we analyze the photometric, spectroscopic, and host-galaxy properties of SN\,2025wiu and compare them with those of other SESNe. In Section~\ref{sec:4}, we model the light curve with \textsc{TransFit} \citep{Liu2025}; in Section~\ref{sec:5}, we discuss the possible powering mechanisms; and Section~\ref{sec:6} summarizes our conclusions. Throughout this paper, we adopt a flat $\Lambda$CDM cosmology with $H_0 = 67.66$~km~s$^{-1}$~Mpc$^{-1}$ and $\Omega_m = 0.310$ (Planck Collaboration et al. \citeyear{planck2020planck}).

\section{Observations and Data Reduction} \label{sec:Observations}

\subsection{Discovery and Basic Properties} \label{subsec:basic}

SN\,2025wiu was discovered by WFST \citep{wang2023science} on 2025 August 28 (MJD = 60915.890) and located at R.A. = 01$^{\rm h}$55$^{\rm m}$58\fs487, decl. = $-$00\arcdeg26\arcmin43\farcs24 (J2000.0). SN\,2025wiu lies $6.72\arcsec$ (6.40~kpc projected) from the nucleus of its host galaxy, KUG\,0153$-$006, whose spectroscopic redshift of $z = 0.046927$ \citep{2017ApJS..233...25A} corresponds to a luminosity distance of $D_L = 215.3$~Mpc and a distance modulus of $\mu = 36.67$~mag under our adopted cosmology.

\subsection{Photometry} \label{subsec:phot}

We obtained point-spread function (PSF) photometry for the ZTF $g$-, $r$- and $i$-band images \citep{masci2019zwicky} and the ATLAS $c$-, $w$- and $o$-band images \citep{smith2020design,2021TNSAN...7....1S}. All photometry was corrected for Milky Way extinction of $E(B-V) = 0.0275$~mag \citep{schlafly2011measuring}, assuming $R_V = 3.1$ \citep{1999PASP..111...63F}. $K$-corrections were derived from synthetic photometry of our spectra and linearly interpolated to the photometric epochs.

\begin{figure*}[ht!]
\centering
\includegraphics[width=\linewidth]{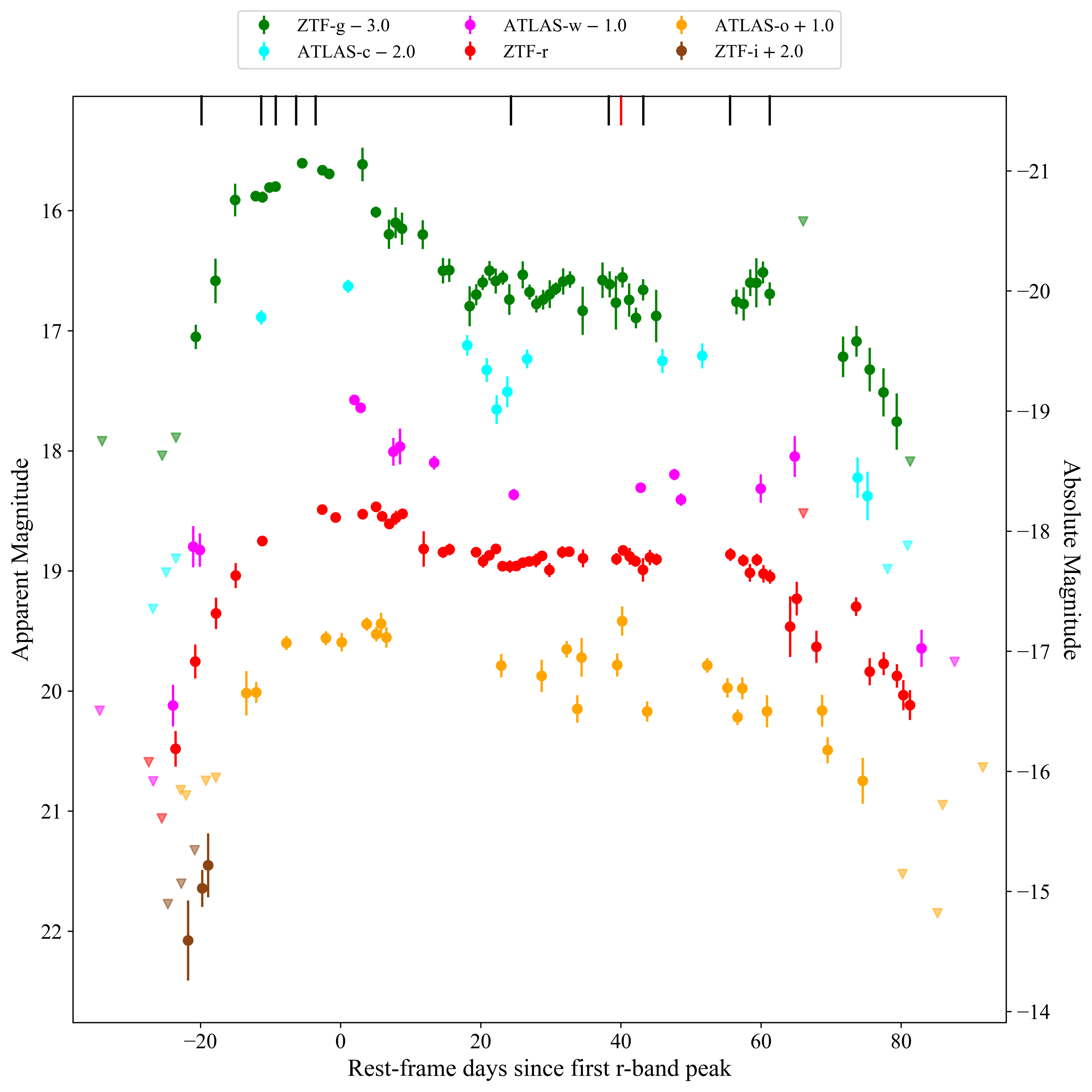}
\caption{Multi-band light curves of SN\,2025wiu in the observed $g$, $r$, $i$, $c$, $w$, and $o$ bands. Downward-pointing triangles indicate $5\sigma$ upper limits. Error bars represent $1\sigma$ photometric uncertainties. The absolute-magnitude scales are corrected for Milky Way extinction and K-correction; no correction for host-galaxy extinction has been applied. Phases are given in the rest frame relative to the first $r$-band maximum (${\rm MJD} = 60945.1$). Vertical tick marks at the top indicate the epochs of spectroscopic observations; red ticks denote near-infrared spectra and black ticks denote optical spectra.
\label{fig:LC}}
\end{figure*}

The complete photometry is shown in Figure~\ref{fig:LC}. All phases are reported in the rest frame after correcting for time dilation. We define phase~$=0$ as the first $r$-band maximum brightness, which corresponds to ${\rm MJD} = 60945.1$. At maximum light, the apparent magnitude of $18.5$~mag corresponds to an absolute magnitude of $-18.2$~mag. The full dataset is provided in Appendix~\ref{sec:A}.

\subsection{Spectroscopy} \label{subsec:spec}

\begin{figure}[ht!]
\centering
\includegraphics[width=\columnwidth]{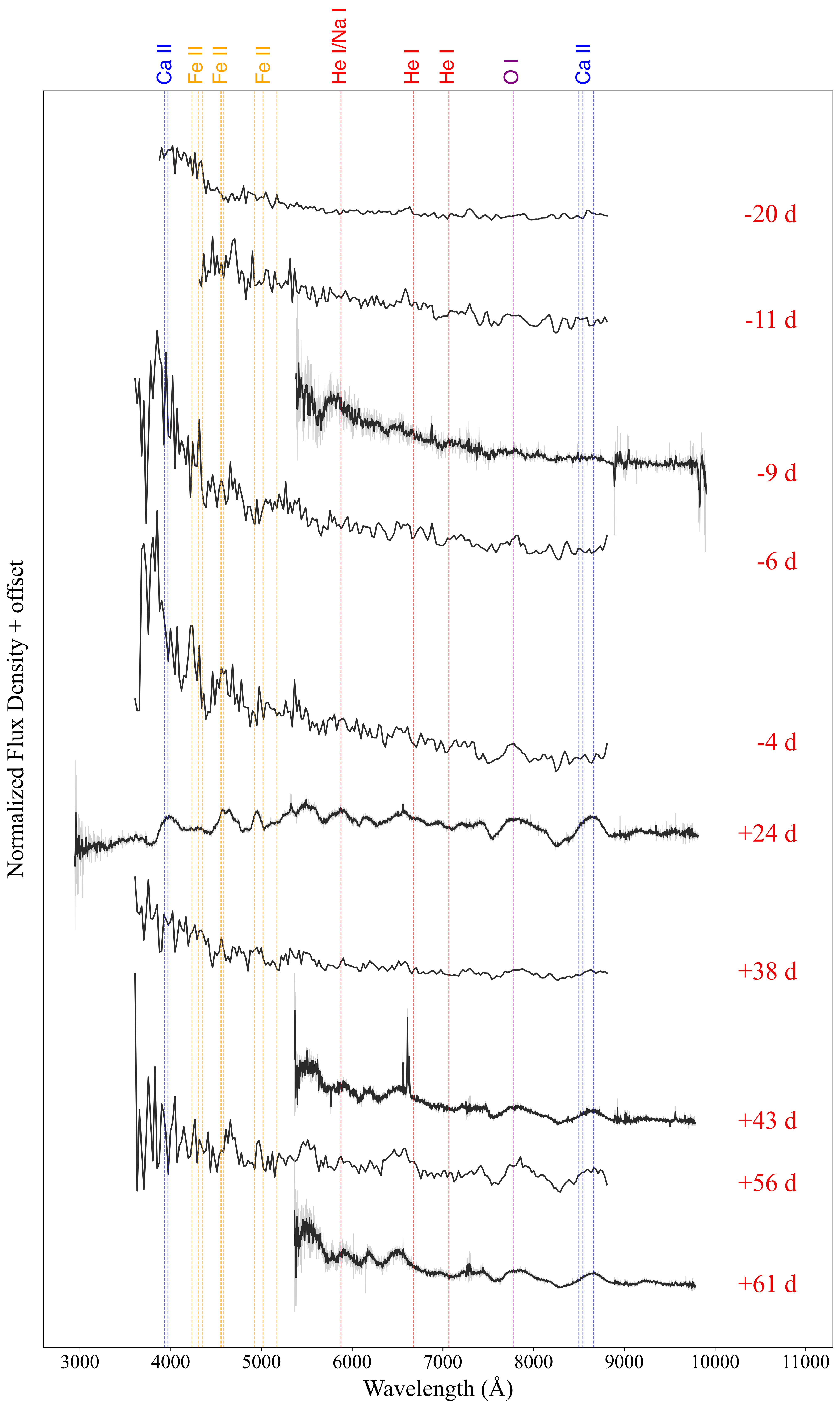}
\caption{Spectral evolution of SN\,2025wiu, shown in the rest frame with the phase of each epoch labeled on the right. The observed spectra are in gray, with smoothed versions overplotted in black. Vertical dashed lines mark prominent features: \ion{Ca}{2} (blue), \ion{Fe}{2} (orange), \ion{He}{1} (red), [\ion{Ca}{2}] (green), and \ion{O}{1} (purple). The prominent \ion{He}{1} lines and absence of hydrogen identify SN\,2025wiu as a SN Ib.
\label{fig:spec}}
\end{figure}

We obtained a total of eleven follow-up spectra spanning phases from $-20$ to $+61$~days, all acquired by the ZTF group. The spectroscopic log is provided in Appendix~\ref{sec:B}. Ten of these are optical spectra: six were taken with the Spectral Energy Distribution Machine \citep[SEDM;][]{2018PASP..130c5003B,2019A&A...627A.115R,2022PASP..134b4505K} on the Palomar 60-inch telescope (P60), three with the Next Generation Palomar Spectrograph (NGPS) on the Palomar 200-inch telescope (P200), and one with the Low Resolution Imaging Spectrometer (LRIS) on the Keck~I telescope. In addition, one near-infrared spectrum was obtained with the Near-Infrared Echellette Spectrometer (NIRES) on the Keck~II telescope. All spectra were retrieved from Fritz, the ZTF science data platform \citep{2019JOSS....4.1247V,2023ApJS..267...31C}. All spectra were corrected for Galactic extinction and transformed to the rest frame by dividing the wavelength axis by $(1+z)$. The optical spectra are shown in Figure~\ref{fig:spec}, and the near-infrared spectrum is shown in Figure~\ref{fig:nir_spec_comparison}. The spectral classification was confirmed using \texttt{SNID-SAGE} \citep{2026MNRAS.549g1066S} and comparison with template Type~Ib spectra. The spectra will be made publicly available through WISeREP \citep{Yaron2012}.

\section{Photometric and Spectroscopic Properties} \label{sec:Photometric}

\subsection{Optical Light-curve Evolution} \label{subsec:lc}

Figure~\ref{fig:LC} shows our light curves. We estimated the epoch of first light ($t_0$) by fitting the early rise with a power law in flux,
\begin{equation}
f_{b}(t) = f_{0,b}\,(t - t_{0})^{n_{b}}, \qquad t > t_{0},
\label{eq:powerlaw}
\end{equation}
and $f_{b} = 0$ before $t_{0}$. Here $f_{b}$ is the Galactic-extinction-corrected flux density in band $b \in \{g, r\}$, $t_{0}$ is shared by both bands, and $f_{0,b}$ and $n_{b}$ are free in each. Following \citet{2020ApJ...902...47M}, we fitted only data below 40\% of the peak flux in each band. These comprise the ZTF forced photometry, including the WFST discovery ($g = 22.74 \pm 0.10$\,mag at MJD~60915.89) reported to TNS. The non-detection enters the likelihood as an upper limit. Because $t_{0}$ and $n_{b}$ are strongly degenerate and the window contains few detections, we quote maximum-likelihood values with 68\% profile-likelihood intervals, which, unlike the marginalized posterior, do not depend on the prior on $n_{b}$. We find $t_{0} = {\rm MJD}\,60913.9^{+1.0}_{-2.3}$, $n_{g} = 1.7^{+0.9}_{-0.5}$, and $n_{r} = 1.9^{+0.8}_{-0.5}$. The result corresponds to a rest-frame rise time to the first peak of $29.8^{+2.3}_{-1.0}$\,days.

\begin{figure*}[ht!]
\centering
\includegraphics[width=\textwidth]{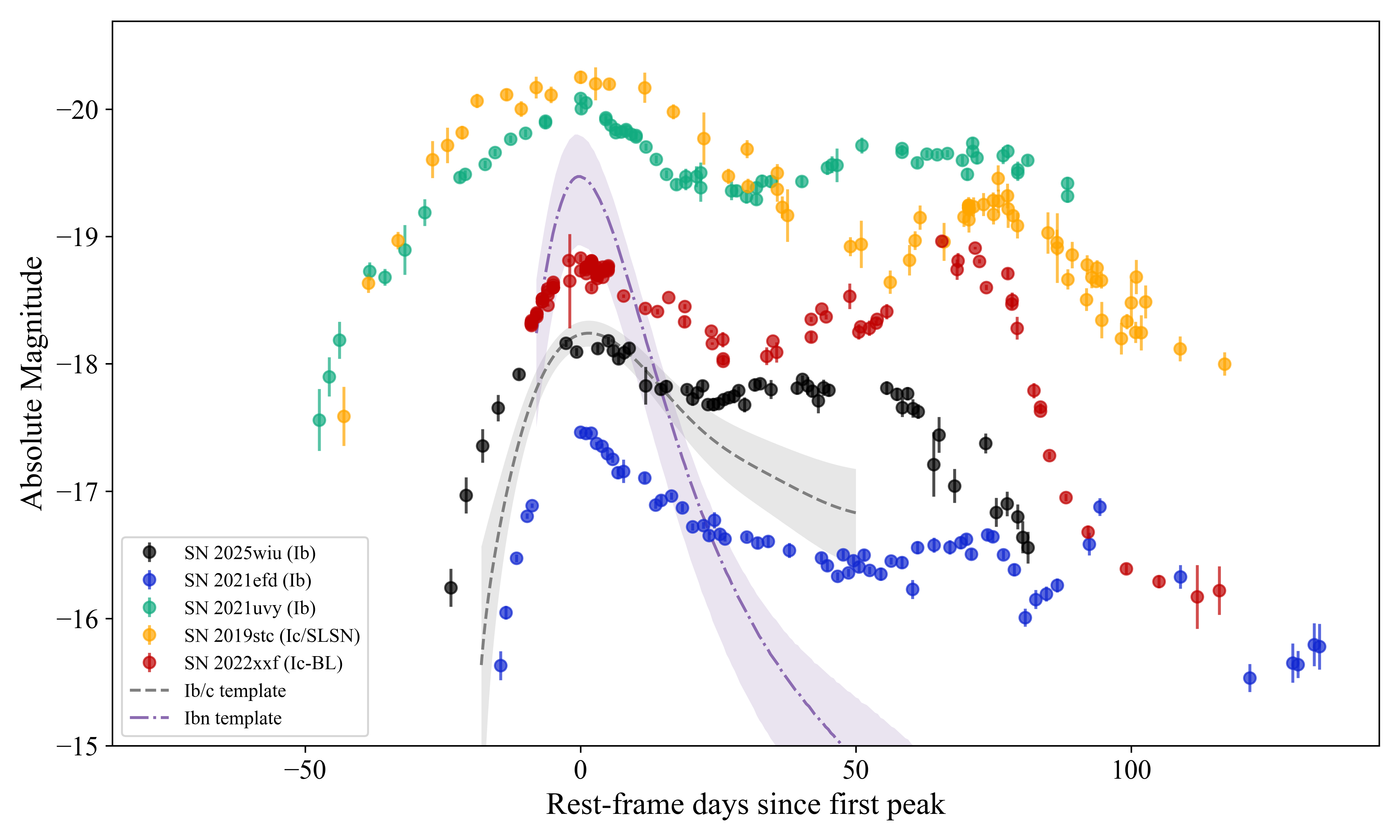}
\caption{$r$-band absolute magnitude light curve of SN\,2025wiu (black) compared with other double-peaked SESNe: SN\,2021efd (blue), SN\,2021uvy (green), SN\,2019stc (orange), and SN\,2022xxf (red). The dashed gray and dash-dotted purple curves show the Type~Ib/c \citep{2015A&A...574A..60T} and Type~Ibn \citep{2017ApJ...836..158H} templates, with shaded regions indicating their $1\sigma$ dispersion. All light curves are corrected for Galactic extinction.
\label{fig:LC_comparison}}
\end{figure*}

At $r$-band maximum, the absolute magnitude reaches $M_r = -18.2$. The subsequent $r$-band decline of $0.023 \pm 0.004$~mag~day$^{-1}$ is comparable to that of normal Type~Ib events. However, after phase=$+15$~days, the luminosity decline in all bands slows and departs from that of typical Type~Ib supernovae (see also Figure~\ref{fig:LC_comparison}), followed by a slight rebrightening after phase=$+23$~days that reaches a second peak at phase=$+38$~days, where $M_r = -17.8$. After phase=$+60$~days, the light curve declines dramatically at a rate of $0.052 \pm 0.016$~mag~day$^{-1}$ in $g$-band and $0.049 \pm 0.007$~mag~day$^{-1}$ in $r$-band.

In Figure~\ref{fig:LC_comparison}, we compare the $r$-band absolute magnitude of SN\,2025wiu with those of other similar double-peaked SESNe, including SN\,2021efd \citep{2026A&A...706A.183P}, SN\,2021uvy \citep{2025PASP..137i4101S}, SN\,2019stc \citep{2021ApJ...913..143G}, and SN\,2022xxf \citep{2023A&A...678A.209K}, together with the Type~Ibc \citep{2015A&A...574A..60T} and Type~Ibn \citep{2017ApJ...836..158H} templates. Like SN\,2025wiu, both SN\,2021uvy and SN\,2019stc exhibit a brighter first peak followed by a slightly fainter second peak, although they are nearly two magnitudes more luminous; their first peaks have been attributed to a combination of magnetar spin-down and $^{56}$Ni radioactive decay \citep{2022ApJ...941..107G, 2025PASP..137i4101S}. The first peak of Type Ic-BL SN\,2022xxf was powered by radioactivity with ejecta-CSM interaction dominating in its second peak \citep{2023A&A...678A.209K}. SN\,2021efd displays normal Type~Ib luminosity at early phases similar to SN\,2025wiu, but its later bump is produced by the ejecta colliding with the clumpy CSM \citep{2026A&A...706A.183P}.

\subsection{Color and Bolometric Evolution} \label{subsec:color&bolo}

\begin{figure}[ht!]
\centering
\includegraphics[width=\columnwidth]{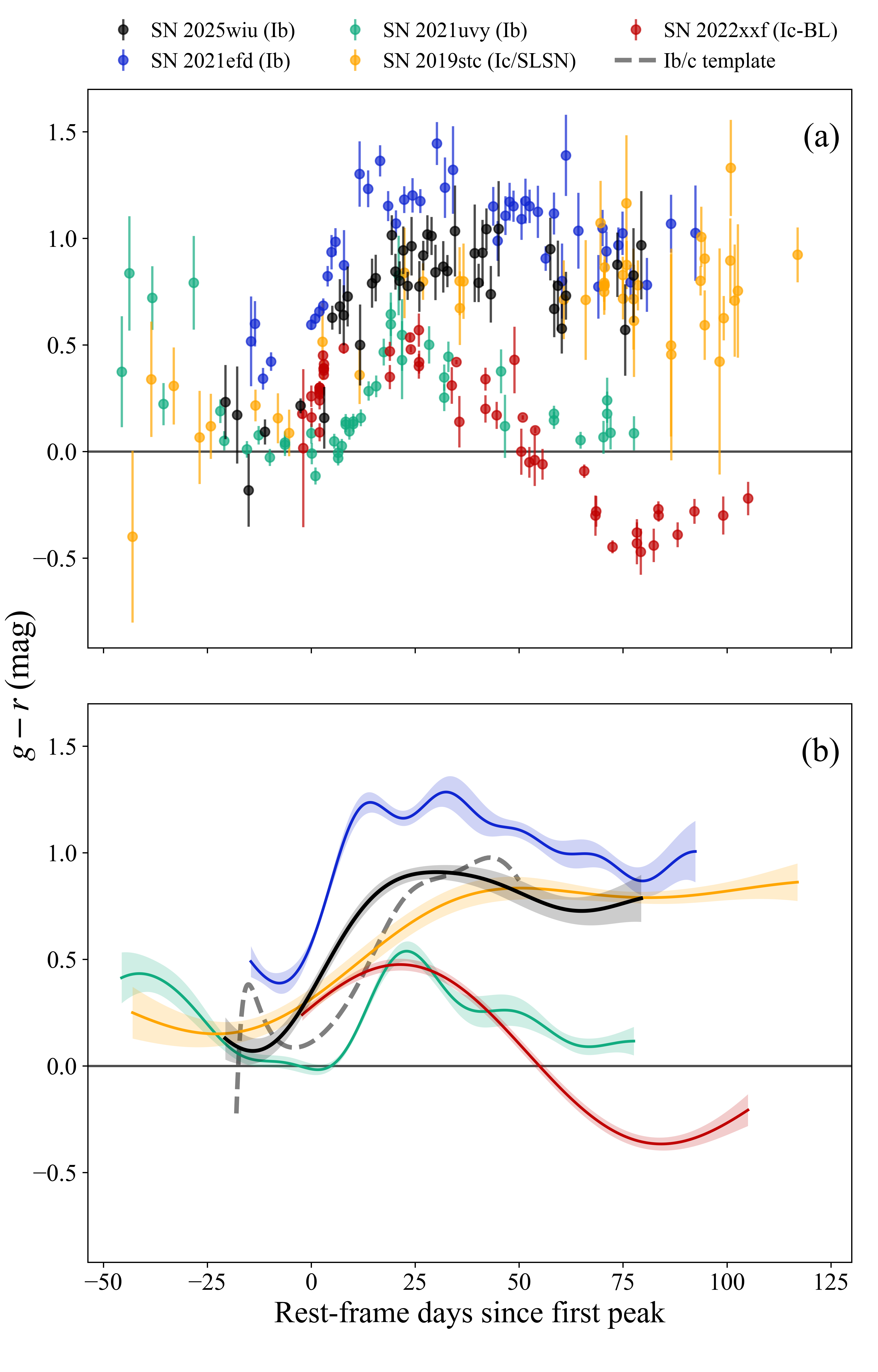}
\caption{Rest-frame $g-r$ color evolution of SN\,2025wiu compared with other double-peaked SESNe. \textbf{(a)}~Individual color measurements with $1\sigma$ error bars for SN\,2025wiu (black), SN\,2021efd (blue), SN\,2021uvy (green), SN\,2019stc (orange), and SN\,2022xxf (red). \textbf{(b)}~Gaussian-process smoothed color curves for the same objects, with shaded regions indicating the $1\sigma$ uncertainty. The dashed gray curve shows the Type~Ibc template from \citet{2015A&A...574A..60T}.
\label{fig:color}}
\end{figure}

Figure~\ref{fig:color} shows the $g-r$ color evolution of SN\,2025wiu compared with other SESNe. At early phases, SN\,2025wiu exhibits a reddening consistent with normal Type~Ibc supernovae, reaching a maximum of $g-r = 0.91 \pm 0.03$~mag at phase=$+29.9$~days. Notably, the color shows no significant blueward evolution even during the second peak, becoming bluest at $+66.0$~days with $g-r = 0.73 \pm 0.06$~mag.

SN\,2025wiu shows relatively weak color evolution during its second light-curve peak compared to SN\,2021uvy and SN\,2022xxf. For SN\,2021uvy, the light-curve bumps have been attributed to magnetar spin-down, which drives a rise in photospheric temperature and a corresponding blueward color evolution \citep{2022MNRAS.513.6210M,2025PASP..137i4101S}. In the case of SN \,2022xxf, the weak narrow lines in the spectra of its second peak, point to the ejecta-CSM interaction \citep{2023A&A...678A.209K}. The color evolution of SN\,2025wiu is similar to that of SN\,2019stc and SN\,2021efd, both of which have been associated with CSM interaction during their second peaks \citep{2021ApJ...913..143G,2025PASP..137i4101S,2026A&A...706A.183P}.

\begin{figure}[ht!]
\centering
\includegraphics[width=\columnwidth]{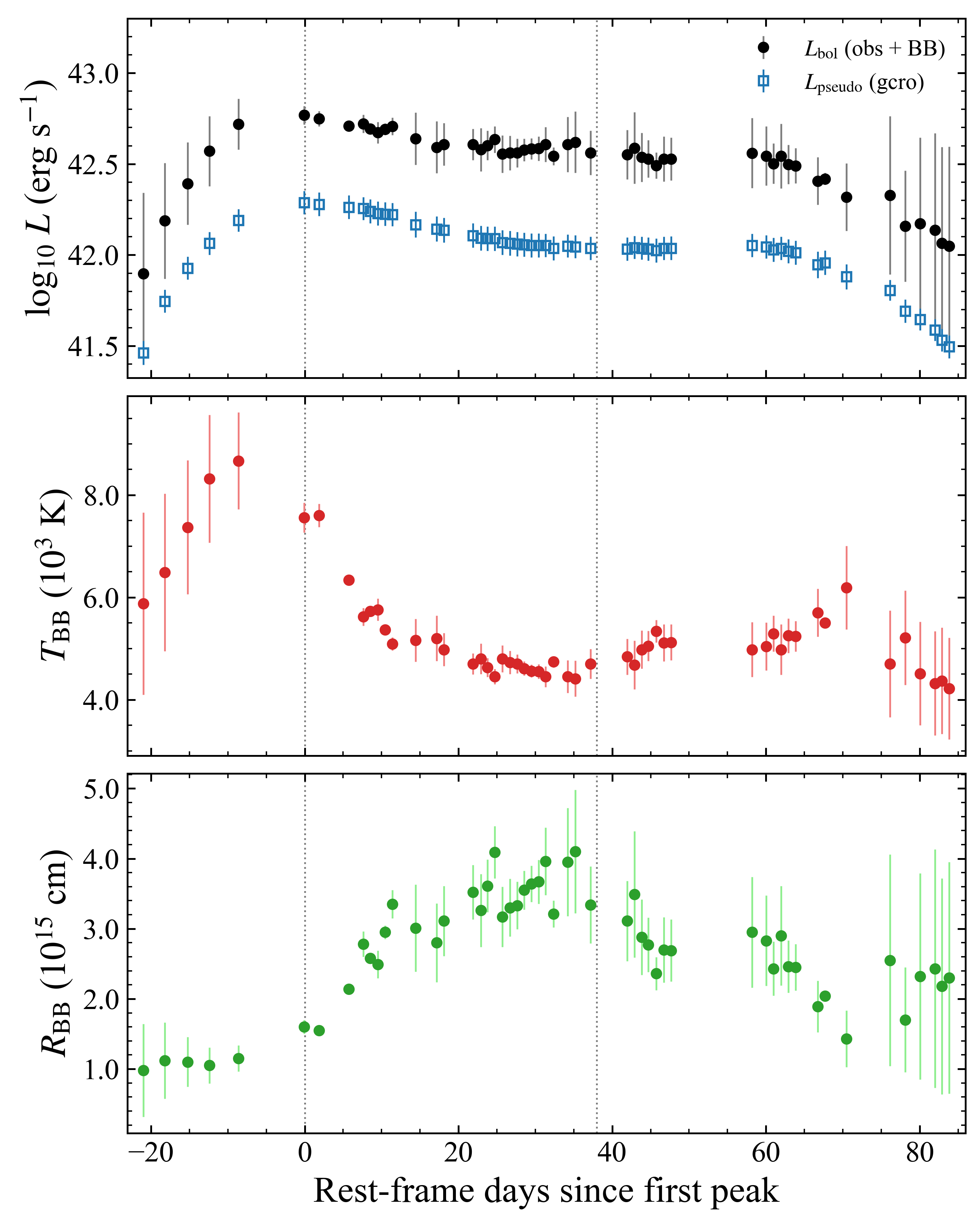}
\caption{Bolometric properties of SN\,2025wiu. \textit{Top:} full bolometric luminosity ($L_{\rm bol}$, black filled circles) and pseudo-bolometric luminosity ($L_{\rm pseudo}$, blue open squares). \textit{Middle:} blackbody temperature $T_{\rm BB}$. \textit{Bottom:} blackbody radius $R_{\rm BB}$. The gray dotted and dashed vertical lines mark the first (phase~$=0$) and second (phase~$\approx 38$\,d) $r$-band maximum, respectively. Error bars represent $1\sigma$ uncertainties.
\label{fig:bol_lc}}
\end{figure}

We construct the bolometric light curve of SN\,2025wiu with \texttt{SuperBol} \citep{2018RNAAS...2..230N} from the extinction- and K-corrected ZTF $g$, $r$ and ATLAS $c$, $o$ photometry (Figure~\ref{fig:bol_lc}). The pseudo-bolometric luminosity, $L_{\rm pseudo}$, is obtained by integrating the observed SED over the rest-frame range $\approx$4370--8080\,\AA. The full bolometric luminosity, $L_{\rm bol}$, additionally accounts for the unobserved ultraviolet and infrared flux by extrapolating a blackbody fit to the $g$, $c$, $r$, and $o$ photometry over 1-25\,000\,\AA; the same fit yields $T_{\rm BB}$ and $R_{\rm BB}$.

The first peak reaches a bolometric luminosity of $L_{\rm bol} = 5.86 \times 10^{42}$~erg~s$^{-1}$. Thereafter, the luminosity declines monotonically; even at the second peak, the luminosity does not increase but only exhibits a reduced decline rate. Notably, after the second peak the bolometric luminosity drops sharply, at a rate of $\sim 0.054 \pm 0.008$~mag\,day$^{-1}$, roughly 5.5 times faster than the $0.0098$~mag\,day$^{-1}$ expected from fully trapped $^{56}$Co decay \citep{1994ApJS...92..527N}. Even when $\gamma$-ray leakage is taken into account, the decline remains too steep to be explained by $^{56}$Co decay \citep{2015MNRAS.450.1295W}. The blackbody temperature peaks at approximately $8700$~K near the first peak, then decreases to a minimum of $\sim 4500$~K at phase=$+35$~days, before rising again to $\sim 5500$~K around the second peak. The blackbody radius rises steadily from $\sim 1 \times 10^{15}$~cm during the first peak to a maximum of $\sim 4 \times 10^{15}$~cm, and begins to decline after phase $+35$~days.

\subsection{Optical Spectral Evolution} \label{subsec:spec_optical}

Figure~\ref{fig:spec} shows the optical spectra of SN\,2025wiu, which broadly display the characteristics of SN Ib. Before the first peak, the spectrum at phase $-20$~days still exhibits a blue continuum. By $-11$ and $-9$~days, \ion{Fe}{2} and \ion{He}{1}~$\lambda 5876$ have emerged, together with weak \ion{Ca}{2} NIR triplet and \ion{O}{1}~$\lambda 7774$ features, and the \ion{He}{1}~$\lambda 5876$ ejecta velocity reaches $\sim 12000$~km\,s$^{-1}$. At $+24$~days the spectrum retains its Type~Ib characteristics, but the pseudo-equivalent width (pEW) of \ion{He}{1} is considerably weaker than in typical Type~Ib events: while the sample of \citet{2023A&A...675A..83H} has a mean \ion{He}{1}~$\lambda 5876$ pEW of $\sim 100$~\AA, we measure only $\sim 36$~\AA. The \ion{He}{1}~$\lambda 6678$ and $\lambda 7065$ lines are barely detectable at this epoch, and the \ion{He}{1}~$\lambda 5876$ velocity has decreased to $\sim 9200$~km\,s$^{-1}$, with \ion{O}{1}~$\lambda 7774$ and \ion{Fe}{2}~$\lambda 5169$ at $\sim 8100$ and $\sim 7700$~km\,s$^{-1}$. Thereafter, until our final spectrum at $+61$~days, the only significant change is a steady decline in velocity, by $\sim 3000$~km\,s$^{-1}$ for \ion{He}{1}~$\lambda 5876$ and $\sim 1000$~km\,s$^{-1}$ for \ion{O}{1}~$\lambda 7774$. The \ion{He}{1}~$\lambda 5876$ pEW increases mildly toward later phases, reaching $50 \pm 9$~\AA\ at $+61$~days, but remains well below typical Type~Ib values throughout. The overall spectral appearance and the luminosity therefore show no substantial change across the second peak.

The measured velocities and pEWs are listed in Tables~\ref{tab:v_pew}. For each line, we defined pseudo-continuum regions on the blue and red sides of the absorption feature, then randomly resampled the pseudo-continuum within these windows and fitted the absorption feature with a Gaussian profile. We repeated this procedure for 1,000 Monte Carlo realizations and report the mean values across all realizations. The pEW uncertainty combines, in quadrature, the uncertainty propagated from the fitted Gaussian amplitude and width and the dispersion of pEW values across realizations, the latter reflecting the sensitivity to pseudo-continuum placement.

\begin{figure*}[!ht]
\centering
\includegraphics[width=\textwidth]{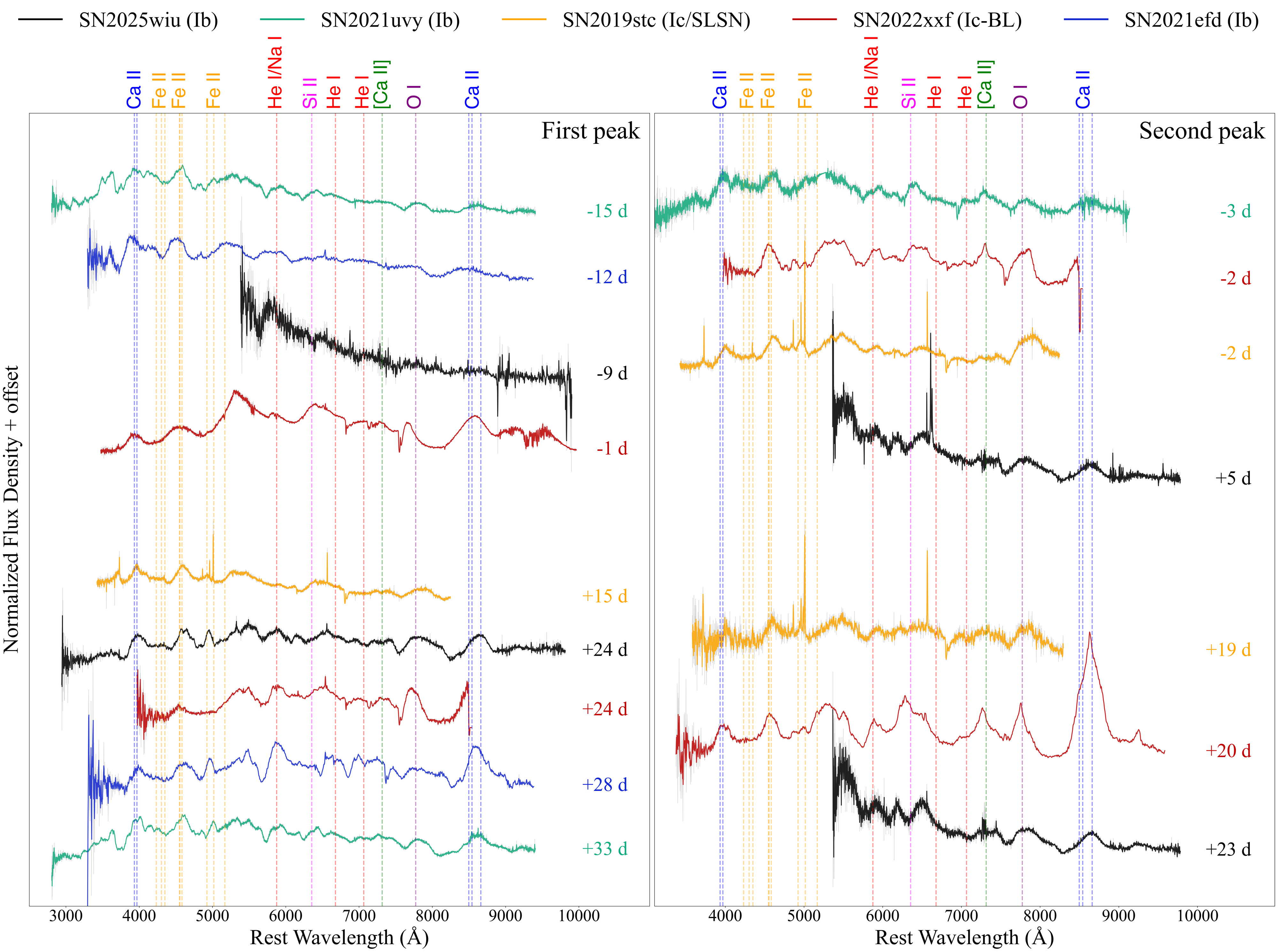}
\caption{Comparison of the spectra of SN\,2025wiu (black) with those of other double-peaked SESNe: SN\,2021uvy (green), SN\,2019stc (orange), SN\,2022xxf (red), and SN\,2021efd (blue). All spectra are shown in the rest frame, normalized and vertically offset, with the phase of each epoch labeled beside the corresponding spectrum. Phases in the left panel are given relative to the first r-band light curve peak, while those in the right panel are relative to each object's own second peak. Vertical dashed lines mark prominent features: \ion{Ca}{2} (blue), \ion{Fe}{2} (orange), \ion{He}{1} (red), \ion{Si}{2} (magenta), [\ion{Ca}{2}] (green), and \ion{O}{1} (purple).
\label{fig:spec_comparison}}
\end{figure*}

Figure~\ref{fig:spec_comparison} (left panel) compares the spectra of SN\,2025wiu with those of the other objects. In the pre-peak phase, SN\,2025wiu, SN\,2021uvy,  and SN\,2021efd all display \ion{He}{1} and \ion{Ca}{2} features, although our spectrum covers only wavelengths longward of $\sim 5400$~\AA\ and therefore does not sample the \ion{Fe}{2} features seen in the other two objects. Between $+24$ and $+33$~days, these three objects again show closely comparable spectra, although the \ion{He}{1} lines of SN\,2021efd are noticeably stronger than in the other two. SN\,2022xxf displays \ion{Fe}{2}, \ion{O}{1}, \ion{Na}{1} and \ion{Si}{2} lines with the broad profiles characteristic of early-phase SNe~Ic-BL \citep{2016ApJ...827...90L}, while SN\,2019stc at $+15$~days shows the typical features of SNe~Ic, including \ion{Ca}{2}~$\lambda\lambda 3934, 3969$, \ion{Fe}{2}, \ion{Na}{1}~$\lambda 5890$, \ion{Si}{2}~$\lambda 6355$, and \ion{O}{1}~$\lambda 7774$ \citep{2021ApJ...913..143G}, with line profiles narrower than those of SN\,2022xxf. Up to this point, none of these objects shows any spectroscopic indication of the second peak that follows.

The right panel of Figure~\ref{fig:spec_comparison} compares the spectra around the second peaks. Before its second peak, SN\,2022xxf shows little emission apart from [\ion{O}{2}]~$\lambda\lambda 7320, 7330$. After the second peak, its spectrum becomes dominated by emission lines, including \ion{Na}{1}~$\lambda 5893$, [\ion{O}{1}]~$\lambda\lambda 6300, 6364$, [\ion{Ca}{2}]~$\lambda\lambda 7292, 7324$, \ion{O}{1}~$\lambda 7774$, and the \ion{Ca}{2} NIR triplet; the narrow components and low velocities ($< 5000$~km\,s$^{-1}$) of these lines have been interpreted as evidence for a CSM origin of the second peak \citep{2023A&A...678A.209K}. In contrast, SN\,2019stc and SN\,2021uvy does not shows significant spectral evolution across its second peak, consistent with the slow spectral evolution of SLSNe \citep{2021ApJ...913..143G,2025PASP..137i4101S}. SN\,2025wiu shows a similar behavior: apart from the gradual decline in line velocities (Tables~\ref{tab:v_pew} ), its spectra remain essentially unchanged from the second peak to the end of our spectroscopic coverage, and no narrow emission components attributable to the supernova are detected. As in SN\,2019stc, the second peak thus leaves no discernible imprint on the spectra, a similarity that is notable given that SN\,2025wiu is $\sim 2$~mag fainter at peak and does not reach SLSN-like luminosities. Even at $+61$~days, the spectrum shows no sign of the forbidden [\ion{O}{1}]~$\lambda\lambda 6300, 6364$ or [\ion{Ca}{2}]~$\lambda\lambda 7292, 7324$ emission that marks the onset of the nebular phase and remains dominated by P-Cygni profiles, indicating that the ejecta stay optically thick throughout the second peak.

\begin{deluxetable*}{ccccccc}
\tablecaption{Ejecta velocities and pseudo-equivalent widths of SN\,2025wiu\label{tab:v_pew}}
\tablehead{
\colhead{Phase} & \multicolumn{4}{c}{Velocity (km\,s$^{-1}$)} & \multicolumn{2}{c}{pEW (\AA)} \\
\colhead{(d)} & \colhead{\ion{He}{1}~$\lambda 5876$} & \colhead{\ion{O}{1}~$\lambda 7774$} &
\colhead{\ion{Fe}{2}~$\lambda 5169$} & \colhead{\ion{Ca}{2}~$\lambda 8542$} &
\colhead{\ion{He}{1}~$\lambda 5876$} & \colhead{\ion{O}{1}~$\lambda 7774$}
}
\startdata
$-9$  & $12110 \pm 330$ & \nodata       & \nodata        & $11811 \pm 453$ & $33.0 \pm 8.5$  & \nodata \\
$+24$ & $9240 \pm 90$   & $8090 \pm 40$ & $7690 \pm 190$ & $10156 \pm 215$ & $36.2 \pm 4.7$  & $72.8 \pm 2.1$ \\
$+43$ & $7220 \pm 130$  & $7510 \pm 50$ & \nodata        & $9973 \pm 334$  & $40.0 \pm 9.7$ & $68.0 \pm 3.3$ \\
$+61$ & $6180 \pm 230$  & $6960 \pm 80$ & \nodata        & $10014 \pm 326$ & $50.2 \pm 8.6$  & $73.7 \pm 6.4$ \\
\enddata
\end{deluxetable*}

\subsection{Near-infrared Spectrum} \label{subsec:spec_nir} 

\begin{figure}[ht!]
\centering
\includegraphics[width=\columnwidth]{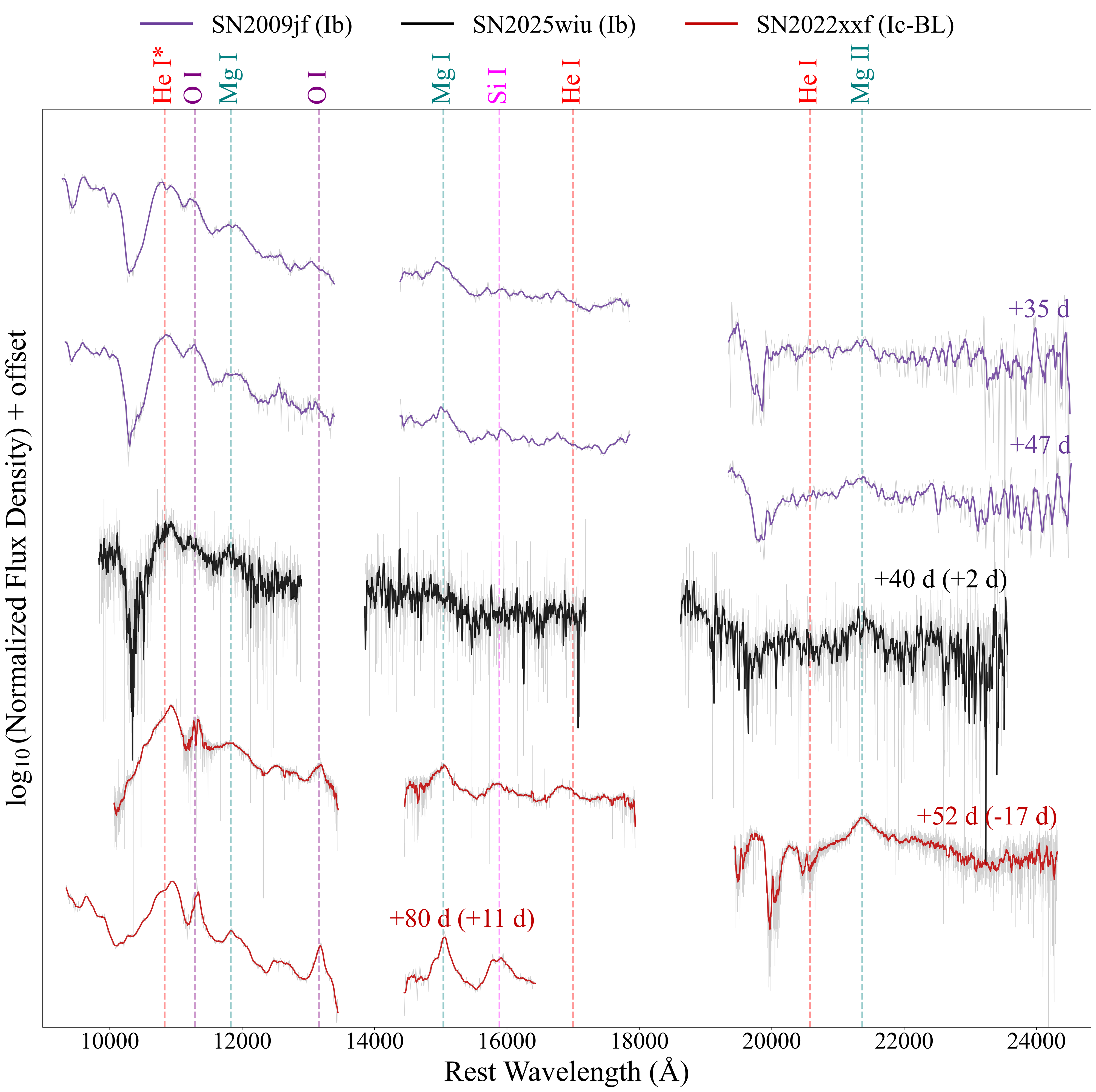}
\caption{Comparison of the near-infrared spectrum of SN\,2025wiu (black) with those of the Type~Ib SN\,2009jf (purple) and the Type~Ic-BL SN\,2022xxf (red). Spectra are shown in the rest frame, with the observed spectra in gray and smoothed versions overplotted. Phases are given relative to the first $r$-band maximum; values in parentheses are relative to the second peak. Gaps correspond to regions of strong telluric absorption. Vertical dashed lines mark prominent features: \ion{He}{1} (red), \ion{O}{1} (purple), \ion{Mg}{1} and \ion{Mg}{2} (teal), and \ion{Si}{1} (magenta). The feature near 1.04--1.08\,\micron, marked \ion{He}{1}$^{*}$, may be a blend of \ion{He}{1}~$\lambda$10830 with \ion{C}{1}~$\lambda$10693, \ion{Mg}{2}~$\lambda$10927, and \ion{Si}{1}~$\lambda$10827; the red dotted lines mark the positions of \ion{C}{1} and \ion{Mg}{2}.
\label{fig:nir_spec_comparison}}
\end{figure}

Figure~\ref{fig:nir_spec_comparison} compares the NIR spectrum of SN\,2025wiu with those of SN\,2022xxf, another double-peaked SESN, and the normal Type~Ib SN\,2009jf \citep{2011MNRAS.416.3138V}. Our NIR spectrum was obtained at $+40$~days, close to the $r$-band second peak.

The spectrum of SN\,2025wiu closely resembles that of SN\,2009jf, exhibiting the \ion{He}{1}~$\lambda 20581$ line characteristic of SN Ib \citep{2002AJ....124..417H, 2009ApJ...702..226M} together with a strong \ion{He}{1}~$\lambda 10830$ blend. Prominent \ion{O}{1}~$\lambda 11290$, \ion{Mg}{1}~$\lambda 11828$, and \ion{Si}{1}~$\lambda 15888$ features are also detected \citep{2011MNRAS.416.3138V,2022ApJ...925..175S}. SN\,2022xxf NIR spectra are dominated by broad emission features and show no clear detection of hydrogen or helium. The feature present near $10800$~\AA~can be attributed to a blend of \ion{C}{1}, \ion{Mg}{2}, and \ion{O}{1}, and the absence of a corresponding \ion{He}{1}~$\lambda 20581$ line argues against a significant helium contribution \citep{2023A&A...678A.209K}. In SN\,2025wiu, by contrast, both \ion{He}{1} lines are clearly detected, confirming its Type~Ib classification. A further distinction lies in the line profiles: SN\,2022xxf develops progressively narrower emission components, consistent with the narrow features seen in its optical spectra and attributed to ejecta-CSM interaction \citep{2023A&A...678A.209K}, whereas the NIR spectrum of SN\,2025wiu around its second peak shows no such narrow components.

\subsection{Host galaxy} \label{subsec:host}

The host galaxy of SN\,2025wiu is KUG\,0153$-$006, an $r = 15.54$~mag galaxy at a spectroscopic redshift of $z = 0.046927$ \citep{2017ApJS..233...25A}. To characterize the host, we assembled ultraviolet-to-infrared imaging from GALEX, Pan-STARRS, and unWISE, and derived matched-aperture photometry across all bands with \textsc{hostphot} \citep{2022JOSS....7.4508M}, adopting an aperture of $11.77'' \times 8.32''$ defined from the Pan-STARRS $r$-band image.

We fitted the resulting SED with \textsc{prospector} \citep{2021ApJS..254...22J}, sampling the posterior with \textsc{dynesty}. The fit adopts the \texttt{Prospector-$\alpha$} model, which incorporates nebular line and continuum emission together with dust emission.

The total stellar mass (\texttt{total\_mass}) is given a log-uniform prior spanning $10^8 < M/M_\odot < 10^{12}$, and the stellar metallicity (\texttt{logzsol}) a top-hat prior over $-2 < \log(Z/Z_\odot) < 0.19$. We adopt a non-parametric star-formation history through the \texttt{z\_fraction} prescription, regularized by a Beta continuity prior that enforces smooth transitions in the star-formation rate across cosmic time. The diffuse dust optical depth (\texttt{dust2}, $\hat{\tau}_2$) is drawn from a uniform prior over $[0, 4.0]$. Infrared dust emission follows the \citet{2007ApJ...657..810D} templates, parameterized by the minimum radiation field intensity ($0.1 < U_{\min} < 25$), the PAH mass fraction ($0.5 < Q_{\rm PAH} < 7.0$), and the mass fraction of dust exposed to intense radiation (\texttt{dust\_gamma}, $\gamma$), for which we assume $0.001 < \gamma < 0.15$. The full posterior distributions of the fitted parameters are shown in Appendix~\ref{sec:C}.

Figure~\ref{fig:host_sed} shows the host-galaxy broadband photometry together with the best-fitting \textsc{prospector} model. The posterior medians and $1\sigma$ uncertainties give a stellar mass of $\log_{10}(M_\ast/M_\odot) = 10.30^{+0.15}_{-0.09}$ and a star-formation rate of ${\rm SFR} = 2.50^{+1.38}_{-1.24}\,M_\odot\,{\rm yr}^{-1}$, corresponding to a specific star-formation rate (sSFR) of $\log({\rm sSFR}) = -9.94^{+0.21}_{-0.25}$\,yr$^{-1}$. The inferred \texttt{dust2} value corresponds to $A_V = 0.235$~mag. Overall, KUG\,0153$-$006 is a typical star-forming galaxy. Relative to the Type~Ib host sample of \citet{2021ApJS..255...29S}, which has median values of $\log_{10}(M_\ast/M_\odot) = 9.92^{+0.13}_{-0.15}$ and $\log({\rm sSFR}) = -9.89^{+0.07}_{-0.07}$, the stellar mass of KUG\,0153$-$006 is higher by $\sim 2.4\sigma$, placing it toward the massive end of the distribution while remaining within the range spanned by Type~Ib hosts. Its sSFR, in contrast, is fully consistent with the sample median.

\begin{figure}[ht!]
\centering
\includegraphics[width=\columnwidth]{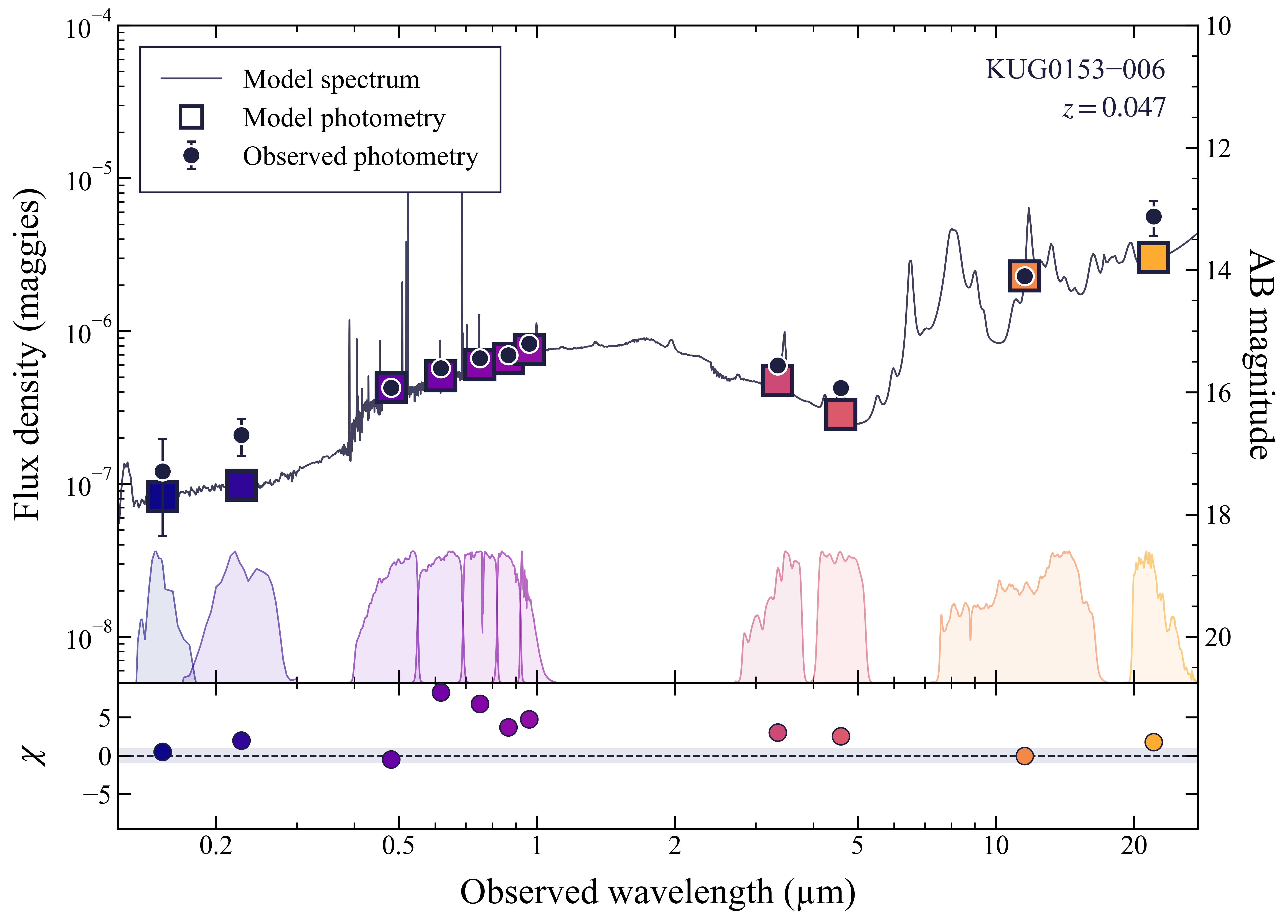}
\caption{Spectral energy distribution of the host galaxy of SN\,2025wiu. \textit{Top:} observed GALEX, Pan-STARRS, and unWISE photometry (circles with $1\sigma$ error bars) compared with the best-fitting \textsc{prospector} model spectrum (solid line) and the corresponding model photometry (open squares). Filter transmission curves are shown at the bottom of the panel. \textit{Bottom:} fit residuals, expressed as $\chi$ in units of the photometric uncertainty. 
\label{fig:host_sed}}
\end{figure}

\section{Light-curve Modeling}
\label{sec:4}

We model the bolometric light curve of
SN~2025wiu to investigate its possible power sources.
As described in Section~\ref{subsec:color&bolo}, the post-peak optical rebrightening
is accompanied by a reduced bolometric decline rate rather than
a distinct second bolometric maximum.
We first fit the early-time light curve with a radioactive-decay
model and then fit the full light curve with a model that includes
ejecta--circumstellar material (CSM) interaction.
These calculations use \texttt{TransFit}
\footnote{\url{https://github.com/CCNUastro/TransFit}}
\citep{Liu2025}, which numerically solves the time-dependent
radiative-diffusion equation in expanding ejecta, and its circumstellar-interaction extension, \texttt{TransFit-CSM} \citep{Zhang2026}, which incorporates
shock heating from ejecta--CSM interaction.

\subsection{ Powering the First Peak}
\label{subsec:4.1}
\begin{figure}
    \centering
    \includegraphics[width=\linewidth]{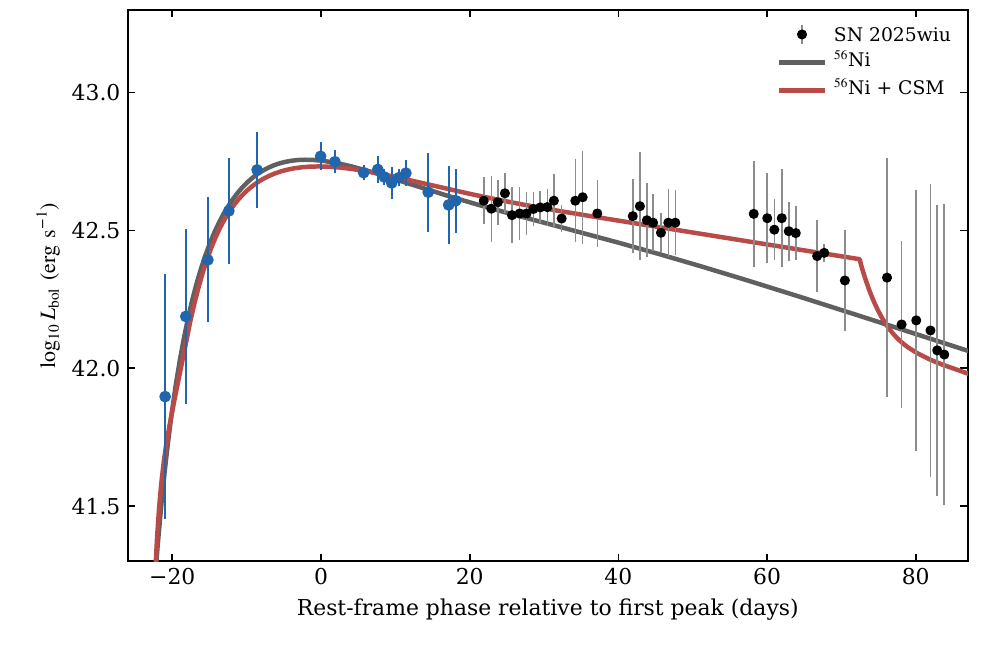}
    \caption{Bolometric light curve of SN~2025wiu
compared with the radioactive-decay model (dark gray curve)
and the combined radioactive-decay and ejecta--CSM interaction
model (red curve).
The blue data points represent the observed data used for fitting the first peak.
(Section~\ref{subsec:color&bolo}).}
    \label{fig:lightcurve_fitting}
\end{figure}

Given the spectroscopic classification of SN~2025wiu as a Type~Ib SN, we first examine whether its first light-curve peak can be powered by the radioactive-decay chain
$^{56}{\rm Ni}\rightarrow{}^{56}{\rm Co}\rightarrow{}^{56}{\rm Fe}$.
Radioactive decay is the standard energy source commonly invoked to explain the main light-curve peak of SESNe. We use the \texttt{nickel} model implemented in TransFit to calculate the deposition and diffusion of radioactive-decay energy in homologously expanding ejecta.

In the calculation, we fix the effective gray $\gamma$-ray opacity to
$\kappa_{\gamma}=0.025~{\rm cm^{2}\,g^{-1}}$,
a value commonly adopted to describe the deposition of radioactive $\gamma$-ray energy in SN ejecta\citep{Swartz1995,Maeda2006}. Based on the expansion velocity inferred from the spectra, we adopt a truncated Gaussian prior for the characteristic ejecta velocity, with a mean of
$7500~{\rm km\,s^{-1}}$
and a standard deviation of
$1500~{\rm km\,s^{-1}}$.

As shown in Figure~\ref{fig:lightcurve_fitting}, the $^{56}$Ni-powered model provides a reasonable description of the overall evolution of the first light-curve peak, with $M_{\rm ej}=5.59^{+3.73}_{-2.25}\,M_{\odot}$, $M_{\rm Ni}=0.47^{+0.06}_{-0.05}\,M_{\odot}$ and $E_k=3.43^{+3.18}_{-1.78}\times10^{51} \rm{erg}\,$. Both values lie at the upper end of those found for normal Type~Ib SNe. The ejecta mass is about twice the mean of $2.6\pm1.1\,M_{\odot}$ reported for this subtype by \citet{Lyman2016} (see also \citealt{Taddia2018,Prentice2019}), although it remains marginally consistent given its large uncertainty; this mainly reflects the long rise time of the first peak ($\approx30$\,d). The $^{56}$Ni mass is substantially higher than the typical values of $\sim0.05$--$0.25\,M_{\odot}$ \citep{Lyman2016,Taddia2018,Prentice2019}. Therefore, although the radioactive-decay model can reproduce the basic morphology of the first peak, it requires both $M_{\rm ej}$ and $M_{\rm Ni}$ at the upper end of the Type~Ib population, suggesting that an additional energy source may contribute to the early-time emission.

\subsection{Modeling the Later Luminosity Excess} \label{subsec:4.2}

To explain the luminosity excess after the first peak, we adopt a combined $^{56}$Ni+CSM model to fit the bolometric luminosity of SN~2025wiu. The CSM density is assumed to follow a power-law distribution, $\rho_{\rm CSM}\propto r^{-s}$, where the density exponent $s$ is a free parameter. Using \texttt{TransFit-CSM}, we compute the shock heating produced by the interaction between the ejecta and the CSM and its subsequent radiative diffusion, and then add the interaction component to the radioactive-decay component. The two components share the ejecta mass, kinetic energy, and explosion time. The optical opacities of the radioactive-decay and CSM components are fixed to $0.1$ and $0.2~\mathrm{cm^2\,g^{-1}}$, respectively, and the characteristic ejecta velocity adopts the same prior as in the pure-nickel model.
The fit results of the combined model are shown in Figure~\ref{fig:lightcurve_fitting}: the inferred ejecta mass is $M_{\rm ej}=4.72^{+2.07}_{-1.59}\,M_\odot$, the kinetic energy is $E_{\rm k}=1.91^{+1.99}_{-1.01}\times10^{51}~\mathrm{erg}$, and the nickel mass is $M_{\rm Ni}=0.378^{+0.070}_{-0.089}\,M_\odot$. The corresponding CSM mass is $M_{\rm CSM}=0.101^{+1.913}_{-0.097}\,M_\odot$, the outer boundary radius is $R_{\rm out}=5.94^{+3.72}_{-3.56}\times10^{5}\,R_\odot$, and the density power-law index is $s=1.95^{+0.60}_{-0.55}$.
In the combined model, CSM interaction contributes about $21.9\%$ of the total model luminosity at the time of the observed first peak and continues to supply additional energy during the subsequent evolution. The combined model indicates that CSM interaction can contribute to the powering of the first peak and provides an explanation for the subsequent luminosity excess.

However, the available spectra of SN~2025wiu do not show prominent arrow emission line features that can be unambiguously attributed to strong CSM interaction. This may indicate that the surrounding CSM is not simply distributed in a spherical geometry\citep{2018MNRAS.477...74A}, but instead has an asymmetric, disk-like, or clumpy structure. Therefore, although the spherically symmetric TransFit-CSM model can reproduce the late-time luminosity excess to some extent, the inferred CSM parameters are more appropriately interpreted as effective quantities characterizing the interaction strength and spatial scale, rather than as unique constraints on the true CSM geometry.

\section{Discussion}
\label{sec:5}
 
Any viable mechanism for the second peak must account for four observations: (i) a second peak at $+38$~days reaching $M_r = -17.8$~mag, accompanied by no increase in the bolometric luminosity, which declines monotonically from $5.9 \times 10^{42}$~erg~s$^{-1}$, but only a reduced decline rate; (ii) a rise in the blackbody temperature from $\sim 4500$ to $\sim 5500$~K coincident with the second peak as $R_{\rm BB}$ recedes, while the $g-r$ color becomes only marginally bluer, by $\sim 0.2$~mag relative to its reddest value; (iii) the absence of narrow or intermediate-width lines in the optical and near-infrared at all epochs, and spectra that remain photospheric from $-20$ to $+61$~days; and (iv) a decline after the second peak of $0.054 \pm 0.008$~mag\,day$^{-1}$, $\sim$5.5 times faster than fully trapped $^{56}$Co decay. We assess each of the mechanisms compiled by \citet{2025PASP..137i4101S} against these constraints.

\subsection{Double-nickel Distribution} \label{subsec:double_ni}
If the second peak were produced by a central $^{56}$Ni component, its epoch would set the diffusion time of the whole ejecta. Solving the model of \citet{1982ApJ...253..785A} with $^{56}$Ni and $^{56}$Co heating, as formulated by \citet{2008MNRAS.383.1485V}, a peak at $\approx68$~d after explosion requires $\tau_m\approx57$~d. Such a model declines by only $\lesssim0.01$~mag\,day$^{-1}$ at 80--110~d, five times slower than observed after the second peak (iv). Adopting the Fe\,\textsc{ii}\,$\lambda5169$ velocity at $+24$~d as the ejecta velocity ($v_{\rm ej}\approx7700$~km\,s$^{-1}$; Table~\ref{tab:v_pew}), it also implies $M_{\rm ej}=\beta c\,v_{\rm ej}\tau_m^{2}/2\kappa\approx10$--$28\,M_\odot$ for $\kappa=0.2$--$0.07$~cm$^{2}$\,g$^{-1}$ and $\beta\simeq13.8$ \citep{1982ApJ...253..785A,2015MNRAS.450.1295W}. This is a lower limit, since any outward mixing of the central $^{56}$Ni would advance the peak, and it exceeds the $6$--$10\,M_\odot$ of published double-nickel models \citep{2018A&A...609A.106T,2021MNRAS.504.4907G,2022A&A...667A..92O}. We therefore disfavour this scenario.

\subsection{Pulsational Pair Instability} \label{subsec:ppi}

In the PPI interpretation of SN\,2020acct \citep{2024ApJ...977L..41A}, the first peak is itself a pre-collapse shell-collision transient, consistent with its flash-ionized narrow He lines and the $\approx2.8$~mag minimum that followed. Neither is seen in SN\,2025wiu: the spectrum at $+24$~d shows broad P-Cygni lines at $\approx7500$--$9000$~km\,s$^{-1}$ typical of SN~Ib photospheres, the first peak is powered mainly by $^{56}$Ni heating with a minor contribution from interaction of the SN ejecta with the CSM (Section~\ref{subsec:4.1}), and the light curve settles onto a slow decline rather than a gap. We therefore associate the first peak with the terminal explosion.

\subsection{Accretion} \label{subsec:accretion}

An alternative energy source is accretion onto the newly formed compact remnant. For SN\,2022jli, it has been proposed that the explosion left a bound compact remnant in a close binary, and that material stripped from the companion and accreted onto the remnant powers the light curve \citep{2024Natur.625..253C,2026A&A...707A.161C}. In this scenario, the orbital motion produces $\approx$12.4-d periodic undulations in the light curve, and the accreted hydrogen gives rise to double-peaked H$\alpha$ and Pa$\beta$ emission whose velocities shift with the same period. SN\,2022jli remains a unique case, and these signatures are specific to the scenario proposed for it rather than generic to accretion-powered SNe. Neither is seen in SN\,2025wiu: the light curve shows no evident periodic modulation, and our spectra at $+38$ and $+61$~d show only broad P-Cygni profiles without such a component. We therefore find no evidence for a SN\,2022jli-like configuration.

\subsection{Magnetar} \label{subsec:magnetar}

\citet{2022MNRAS.513.6210M} showed that a temporary increase in the thermalization efficiency of magnetar spin-down power produces a light-curve bump accompanied by a rise in temperature at nearly constant photospheric radius, qualitatively similar to (ii). In their models, however, the magnetar already powers the main light curve, and the bumps arise from a 2--3-fold increase in the thermalization efficiency lasting 5--20~d; the mechanism was proposed for luminous SNe such as SN\,2019stc. SN\,2025wiu is considerably fainter, and its first peak is powered mainly by $^{56}$Ni (Section~\ref{subsec:4.1}), leaving little room for a magnetar contribution. Because the spin-down power declines monotonically and its thermalized fraction cannot exceed unity, such a weak magnetar would have even less power available at the second peak. Powering the second peak would therefore require the thermalization efficiency to rise by a much larger factor, and over the $\approx43$-d slow phase rather than 5--20~d, at a time when the ejecta are becoming increasingly transparent to high-energy photons. We therefore consider a magnetar unlikely to power the second peak.

\subsection{CSM} \label{subsec:csm}

Our adopted $^{56}$Ni+CSM model (Section~\ref{subsec:4.2}) reproduces the full light curve, with $M_{\rm Ni} \approx 0.38\,M_\odot$, $M_{\rm ej} \approx 4.7\,M_\odot$, $E_{\rm k} \approx 1.9 \times 10^{51}$~erg, and an effective $M_{\rm CSM} \approx 0.10\,M_\odot$ extending to $\sim 5.9\times10^{5}\,R_\odot$, with the interaction contributing from the first peak onward. The subsequent rapid decline of $0.054$~mag\,day$^{-1}$ is as expected once the interaction input terminates \citep{2012ApJ...746..121C}. The absence of narrow lines does not rule out interaction \citep{2012ApJ...747..118M,2013ApJ...773...76C,2018MNRAS.477...74A}; for example, the Type~Ib SN\,2019tsf rebrightened $\approx70$~d after discovery without narrow lines \citep{2020A&A...643A..79S}.

Narrow lines are absent if the pre-shock CSM is either fully ionized, by a high X-ray luminosity or at low density, or insufficiently ionized by the shock \citep{2013ApJ...773...76C}, which we cannot test without X-ray observations. UV and radio data, also unavailable, would offer independent tests: the interaction emerges mainly in the UV \citep{2022A&A...660L...9D}, where narrow \ion{C}{3}, \ion{C}{4}, and \ion{Mg}{2} lines can outlast their optical counterparts \citep{2022ApJ...927..180P}, while its synchrotron emission in the radio traces the CSM density \citep{1998ApJ...499..810C,2006ApJ...651..381C}. They are also suppressed if an optically thick CSM broadens them by electron scattering \citep{2001MNRAS.326.1448C,2018MNRAS.475.1261H}, but the photosphere would then lie in the CSM, hiding the ejecta features \citep{2012ApJ...746..121C} and diluting them with continuum interaction \citep{2015A&A...574A..61L}. Neither is seen in SN~2025wiu: its optical and near-infrared spectra show the broad P-Cygni profiles of the ejecta throughout, with ion-dependent velocities typical of SNe~Ib; the expansion velocity implied by the maximum blackbody radius ($\approx7300$~km\,s$^{-1}$) matches the Fe\,\textsc{ii} velocity ($7690$~km\,s$^{-1}$), placing the photosphere within the ejecta; and the O\,\textsc{i} pEW remains nearly constant from $+24$ to $+43$~d (ratio $0.93\pm0.05$).

We instead favour an aspherical CSM, such as a disk, engulfed by the ejecta, so that the interaction occurs below the ejecta photosphere and its luminosity emerges thermalized, without narrow lines \citep{2015MNRAS.449.1876S,2018MNRAS.477...74A}. In two-dimensional radiation-hydrodynamic simulations, the ejecta progressively cover such a disk, which remains hidden until the ejecta become largely transparent, and an observer near the symmetry axis sees the ejecta directly \citep{2019ApJ...887..249S}. Its spectra only show the broad P-Cygni profiles of a normal SN~Ib, without narrow or intermediate-width components. During the second peak, $R_{\rm BB}$ keeps receding while the temperature rises at nearly constant luminosity, so the photosphere stays within the ejecta and the energy is deposited below it. Our line of sight therefore does not pass through the disk, and we observe the ejecta photosphere and a normal SN~Ib spectrum. The spherically symmetric CSM parameters in Section~\ref{subsec:4.2} should be regarded as effective values describing the energy required by the interaction, rather than the actual CSM geometry.

\section{Conclusions} \label{sec:6}

We have presented optical photometry and optical/near-infrared spectroscopy of SN\,2025wiu, a Type~Ib supernova whose optical light curves show a second maximum. The $r$-band light curve rises for $29.8^{+2.3}_{-1.0}$~days to $M_r = -18.2$~mag and rebrightens after $+23$~days. Its spectra, however, remain photospheric and typical of a normal SN~Ib at all epochs. Our adopted $^{56}$Ni+CSM model, with the interaction contributing from the first peak onward, gives $M_{\rm Ni} \approx 0.38\,M_\odot$, $M_{\rm ej} \approx 4.7\,M_\odot$, and an effective $M_{\rm CSM} \approx 0.10\,M_\odot$.

Of the mechanisms considered, we favour interaction with an aspherical CSM, such as a disk, engulfed by the ejecta. The interaction then occurs below the ejecta photosphere and its luminosity emerges thermalized, without narrow lines; with our line of sight not passing through the disk, we see only the ejecta, consistent with the normal SN~Ib spectra and the photosphere receding within the ejecta during the second peak. 

For similar events, late-time spectroscopy offers the most direct test: once the ejecta become optically thin, the interaction region around an embedded disk should become exposed \citep{2019ApJ...887..249S}, and narrow or intermediate-width emission hidden at earlier phases may then appear. Late-time X-ray and radio observations would further constrain the density and location of the CSM.

\begin{acknowledgments}
We thank Shubham Srivastav for assistance with the ATLAS forced photometry. 
We also thank Takashi Moriya for helpful suggestions.
C.-H. L., T.-W.C. and A.A. acknowledge financial support from the Yushan Fellow Program by the Ministry of Education, Taiwan (MOE-111-YSFMS-0008-001-P1), and the National Science and Technology Council, Taiwan (NSTC grant 114-2112-M-008-021-MY3).
S.Y. acknowledges funding from the National Natural Science Foundation of China under Grant No. 12303046, the Startup Research Fund of Henan Academy of Sciences under Grant No. 242041217, and the Joint Fund of Henan Province Science and Technology R\&D Program under Grant No. 235200810057.
This work has made use of data from the Asteroid Terrestrial-impact Last Alert System (ATLAS) project. The Asteroid Terrestrial-impact Last Alert System (ATLAS) project is primarily funded to search for near earth asteroids through NASA grants NN12AR55G, 80NSSC18K0284, and 80NSSC18K1575; byproducts of the NEO search include images and catalogs from the survey area. The ATLAS science products have been made possible through the contributions of the University of Hawaii Institute for Astronomy, the Queen’s University Belfast, the Space Telescope Science Institute, the South African Astronomical Observatory, and The Millennium Institute of Astrophysics (MAS), Chile.
Based on observations obtained with the Samuel Oschin Telescope 48-inch and the 60-inch Telescope at the Palomar Observatory as part of the Zwicky Transient Facility project. ZTF is supported by the National Science Foundation under Award \#2407588 and a partnership including Caltech, USA; Caltech/IPAC, USA; University of Maryland, USA; University of California, Berkeley, USA; Cornell University, USA; Drexel University, USA; University of North Carolina at Chapel Hill, USA; Institute of Science and Technology, Austria; National Central University, Taiwan, and the German Center for Astrophysics (DZA), Germany. Operations are conducted by Caltech's Optical Observatory (COO), Caltech/IPAC, and the University of Washington at Seattle, USA. SED Machine is based upon work supported by the National Science Foundation under Grant No.~1106171. The Gordon and Betty Moore Foundation, through both the Data-Driven Investigator Program and a dedicated grant, provided critical funding for SkyPortal. The ZTF forced-photometry service was funded under the Heising-Simons Foundation grant \#12540303 (PI: Graham).
We used Claude (Anthropic) for grammar checking and for suggesting language improvements to the manuscript text. The authors reviewed all suggestions and take full responsibility for the content of this paper.

\end{acknowledgments}

\facilities{WFST:2.5m, PO:1.2m (ZTF), ATLAS, PO:1.5m (SEDM), Hale (NGPS), Keck:I (LRIS), Keck:II (NIRES), GALEX, PS1, WISE, NEOWISE}

\software{
    {TransFit} \citep{Liu2025},
    {TransFit-CSM} \citep{Zhang2026},
    astropy \citep{2013A&A...558A..33A,2018AJ....156..123A,2022ApJ...935..167A},
    scipy \citep{2020SciPy-NMeth}, 
    numpy \citep{2020Natur.585..357H},       
    matplotlib \citep{Hunter:2007},
    AutoPhOT \citep{brennan2022automated},
    emcee \citep{2013PASP..125..306F},
    SuperBol \citep{2018RNAAS...2..230N},
    SNID-SAGE \citep{2026MNRAS.549g1066S},
    HostPhot \citep{2022JOSS....7.4508M},
    Prospector \citep{2021ApJS..254...22J},
    dynesty \citep{2020MNRAS.493.3132S},
    FSPS \citep{conroy2009propagation, conroy2010fsps}
    }

\clearpage
\appendix

\section{Photometric Data}\label{sec:A}

The full photometry of SN\,2025wiu is available in machine-readable form.

\section{Spectroscopy data} \label{sec:B}

The spectroscopic observations of SN\,2025wiu are summarized in Table~\ref{tab:spec_log}.

\startlongtable
\begin{deluxetable*}{lccccc}
\tablecaption{Log of spectroscopic observations of SN\,2025wiu\label{tab:spec_log}}
\tablehead{
\colhead{UT Date} & \colhead{MJD} & \colhead{Phase} &
\colhead{Telescope/Instrument} & \colhead{Exposure} &
\colhead{Wavelength Range} \\
\colhead{} & \colhead{} & \colhead{(d)} & \colhead{} & \colhead{(s)} &
\colhead{(\AA)}
}
\startdata
2025-09-06 & 60924    & $-19.88$ & P60/SEDM       & 2700    & 3777--9223  \\
2025-09-15 & 60933    & $-11.32$  & P60/SEDM       & 2700    & 3777--9223  \\
2025-09-17 & 60935 & $-9.26$  & P200/NGPS      & 300     & 5620--10370 \\
2025-09-20 & 60938    & $-6.35$  & P60/SEDM       & 2700    & 3777--9223  \\
2025-09-23 & 60941    & $-3.59$  & P60/SEDM       & 2700    & 3777--9223  \\
2025-10-22 & 60970 & $+24.30$ & Keck\,I/LRIS   & 600     & 3082--10276 \\
2025-11-06 & 60985    & $+38.26$ & P60/SEDM       & 2700    & 3777--9223  \\
2025-11-08 & 60987    & $+40.02$ & Keck\,II/NIRES & \nodata & 9651--24670 \\
2025-11-11 & 60990 & $+43.19$ & P200/NGPS      & 900     & 5580--10375 \\
2025-11-24 & 61003    & $+55.57$ & P60/SEDM       & 600     & 3777--9223 \\
2025-11-30 & 61009 & $+61.24$ & P200/NGPS      & 600     & 5619--10244 \\
\enddata
\tablecomments{Phases are given in the rest frame relative to the first $r$-band maximum (MJD~60945.1)}
\end{deluxetable*}

\clearpage
\section{Host-Galaxy SED Fitting Posteriors} \label{sec:C}

\begin{figure*}[h!]
\centering
\includegraphics[width=\textwidth]{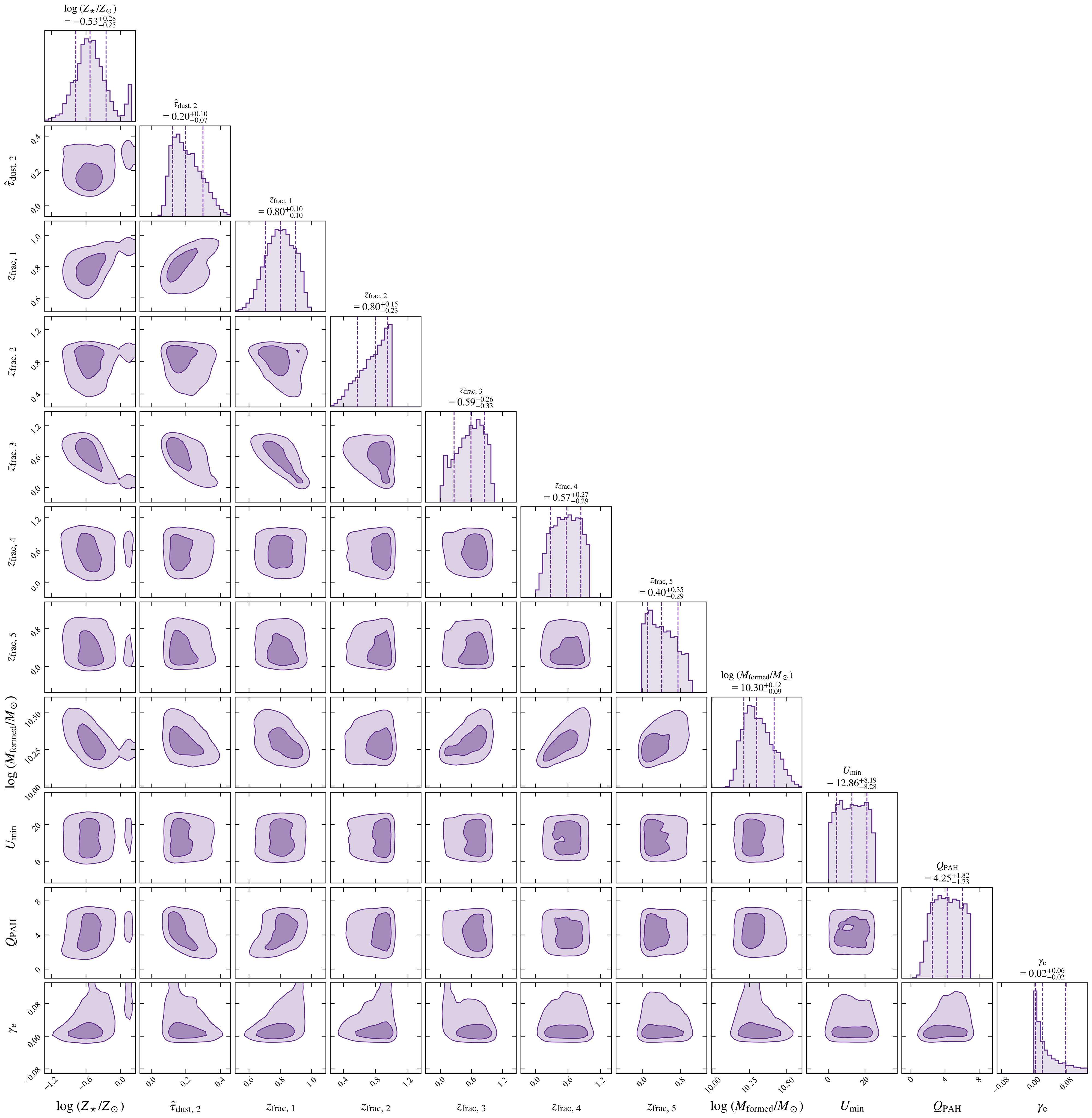}
\caption{Posterior distributions of the \textsc{prospector} model parameters for the host galaxy of SN\,2025wiu. Contours enclose the $1\sigma$ and $2\sigma$ credible regions, and dashed lines in the marginalized histograms mark the 16th, 50th, and 84th percentiles. Quoted values are the posterior medians with $1\sigma$ uncertainties.
\label{fig:host_corner}}
\end{figure*}

Figure~\ref{fig:host_corner} shows the posterior distributions of the free parameters in the \textsc{prospector} fit to the host-galaxy SED.

\clearpage
\bibliography{main}{}
\bibliographystyle{aasjournalv7}

\end{document}